\documentclass[12pt]{article}
\usepackage{amssymb,amsmath,epsfig}
\usepackage{lscape,color,float}
\usepackage{epstopdf}
\begin{document}

\title{\bf Thermodynamics in $f(R,R_{\alpha\beta}R^{\alpha\beta},\phi)$ theory
of gravity}
\author{M. Zubair$^{a,}$\thanks{mzubairkk@gmail.com; drmzubair@ciitlahore.edu.pk},
Farzana Kousar$^{a,}$\thanks{farzana.kousar83@gmail.com} and Sebastian
Bahamonde$^{b,}$\thanks{sebastian.beltran.14@gmail.com}\\ \\
$^{a}$Department of Mathematics, COMSATS\\
Institute of Information Technology Lahore, Pakistan.\\\\
$^{b}$Department of Mathematics, University College London,\\
Gower Street, London, WC1E 6BT, United Kingdom}

\date{}

\maketitle

\begin{abstract}

First and second laws of black hole thermodynamics are examined at the
apparent horizon of FRW spacetime in $f(R,R_{\alpha\beta}R^{\alpha\beta}
,\phi)$ gravity, where $R$, $R_{\alpha\beta}R^{\alpha\beta}$ and $\phi$
are the Ricci scalar, Ricci invariant and the scalar field respectively.
In this modified theory, Friedmann equations are formulated for any spatial
curvature. These equations can be presented into the form of first law of
thermodynamics for $T_{h}d\hat{S}_{h}+ T_{h}d_{i}\hat{S}_{h}+W dV=dE$, where
$d_{i}\hat{S}_{h}$ is an extra entropy term because of the non-equilibrium
presentation of the equations and $T_{h}d\hat{S}_{h}+W dV=dE$ for the equilibrium
presentation. The generalized second law of thermodynamics (GSLT) is expressed in an
inclusive form where these results can be represented in GR $f(R)$ and
$f(R, \phi)$ gravities. Finally to check the validity of GSLT, we take some
particular models and produce constraints of the parameters.
\end{abstract}

\section{Introduction}

The cosmological observations which leads to the recent accelerating
expansion of the universe are weak lensing \cite{1}, large scale structure
\cite{2}, cosmic microwave background (CMB) radiation \cite{3,4}, Type Ia
Supernovae \cite{5} and baryon acoustic oscillations \cite{6}. To explain the
cosmic acceleration of the universe, two classical approaches are followed:
first is to use General Relativity (GR) and introduce "dark energy"
\cite{7,8}; and the second is modified theories of gravity, e.g., $f(R)$
gravity. The modified gravity theories have attain a great attention to study
the current cosmic acceleration \cite{3}.

The first studies in black hole thermodynamics was done in 1970s, when
physicists were thinking that there must be some connection among Einstein
equations and thermodynamics because of the linkage between horizon area
(geometric quantity) and entropy (thermodynamical quantity) of black holes.
At that time, those thermodynamic studies were focused in the context of black
hole in which the surface gravity (geometric quantity) is associated with its
temperature (thermodynamical quantity)  and the first law of thermodynamics
(FLT) is satisfied by these quantities \cite{9}. By the discovery of black
hole thermodynamics, it was shown that gravitation and thermodynamics are
deeply connected \cite{14}. The Hawking temperature of apparent horizon
(defined in proportionality relation with $K_{sg}$ surface gravity) and
horizon entropy $S=A/4G$ fulfil the first law of thermodynamics \cite{9,17}.
In 1995, using that the entropy is proportional to the horizon area of the BH
and the first law of thermodynamics $\delta Q = T dS$, Jacobson \cite{10} was
able to derive the Einstein's equations.
For radiation dominated (FRW) universe, Verlinde discovered that the
Friedmann equation can be recomposed in the form like the Cardy-Verlinde
formula \cite{11}. In higher-dimensional spacetime, this formula represents
an entropy relation for a conformal field theory. It can be observed that
radiation can be described by a conformal field theory. Therefore,
thermodynamics of radiation in the universe has been formulated with the help
of entropy formula can also be write in the form of Friedmann equation, which
describes the dynamics of spacetime. Further, Verlinde discovered the
relation between thermodynamics and Einstein's equations. The discussion
related to the relation between thermodynamics and the Einstein's equation
was done in \cite{13}.

The connection between the first law of thermodynamics (FLT) and the field
equations in Einstein and modified gravities has been extensively studied in
literature. Padmanabhan \cite{13*} developed such connection in Einstein
gravity for spherically symmetric BH and showed that the field equations can
be expressed in the form of FLT, $dE + PdV = TdS$. Such study is also
executed in Lanczos-Lovelock gravity for spherically symmetric and general
static spacetimes \cite{13**}. Applying the first law of thermodynamics to
the apparent horizon of FRW universe and assuming the geometric entropy given
by a quarter of the apparent horizon area, Cai and Kim derived the Friedmann
equations which describes the dynamics of universe with any spatial curvature
\cite{24}. The relation between the Friedmann equations with the first law of
thermodynamics for scalar-tensor gravity and $f(R)$ gravity is discussed by
Akbar \cite{25}. To satisfy the GSLT constraints and conditions imposed on
cosmological future horizon $R_{h}$, Hubble parameter $H$, and the
temperature $T$, in a phantom-dominated universe are described in \cite{26}.
Akbar \cite{28} has shown that the differential form of Friedmann equations
of FRW universe filled with a viscous fluid can be rewritten as a similar
form of the first law of thermodynamics at the apparent horizon of FRW
universe. In \cite{28**}, authors discussed the existence of
Reissner-Nordström and Kerr-Newman black holes in $f(R)$ theories and also
explored their thermodynamics properties $f(R)$ theories and extended
electromagnetic theories. Bamba has studied the first and second laws of
thermodynamics of the apparent horizon in $f(R)$ gravity in the Palatini
formalism \cite{29,30}. He also explored both nonequilibrium and equilibrium
descriptions of thermodynamics in $f(R)$ gravity and conclude that
equilibrium framework is more transparent than the non-equilibrium one. In
\cite{31} the laws of thermodynamics are studied by Wu for the generalized
$f(R)$ gravity with curvature–matter coupling in spatially homogeneous,
isotropic FRW universe whose results shows that the field equations of the
generalized $f(R)$ gravity with curvature–matter coupling can be cast to the
form of the first law of thermodynamics with the the entropy production terms
and the GSLT can be given by considering the FRW universe filled only with
ordinary matter enclosed by the dynamical apparent horizon with the Hawking
temperature.

Further, the laws of thermodynamics at the apparent horizon of FRW spacetime
in modified gravities involving non-minimal matter geometry coupling are
discussed in \cite{22}-\cite{23}. In $f(R,T)$ and $f(R,T,Q)$ gravities it was
found that the picture of equilibrium thermodynamics is not feasible in these
theories, so the non-equilibrium treatment is used to study the laws of
thermodynamics in both forms of the energy-momentum tensor of dark
components. Recently, Huang et al. \cite{32}, presented the thermodynamic
laws for the scalar-tensor theory with non-minimally derivative coupling.

In this paper, the horizon entropy is constructed from the first law of
thermodynamics corresponding to the Friedmann equations in the context of
$f(R,R_{\alpha\beta}R^{\alpha\beta},\phi)$ gravity. We explore the
generalized second law of thermodynamics (GSLT) and find out the necessary
condition for its validity in preview of some well known models. The paper is
organized as follows: In Sec.~\ref{sec2}, we review
$f(R,R_{\alpha\beta}R^{\alpha\beta},\phi)$ gravity and formulate the field
equations of FRW universe. Sec. \ref{sec3} is devoted to study the
non-equilibrium description of first and second laws of thermodynamics. In
Sec. \ref{sec4}, the validity of GSLT for different models are discussed.
Sec. \ref{sec5} is devoted to study the equilibrium description of first and
second laws of thermodynamics. Finally in Sec. \ref{sec6} we conclude our
results. Throughout the paper we will use the metric signature $(-,+,+,+)$,
$c=1$, $\kappa=8\pi G$ and that the Ricci tensor
$R_{\mu\nu}=R^{\sigma}{}_{\mu\sigma\nu}$.

\section{$f(R, R_{\alpha\beta}R^{\alpha\beta},\phi)$ gravity}\label{sec2}

Scalar tensor modified theories of gravity which are based on non-minimal
coupling between matter and the geometry, have had very interesting
applications in the thermodynamics context (See for instance \cite{32*}). Let
us consider a generic theory based on a smooth arbitrary function $f(R,
R_{\alpha\beta}R^{\alpha\beta},\phi)$ on its arguments, where $R$,
$R_{\alpha\beta}R^{\alpha\beta}\equiv Y$ and $\phi$ are the Ricci scalar, the
Ricci invariant and the scalar field respectively within a scalar tensor
context. The action of this modified theory reads \cite{33},
\begin{equation}\label{2.1}
S_{m}=\int d^{4}x \sqrt{-g} \left[\frac{1}{\kappa}\left(f\left(R,R_{\alpha
\beta} R^{\alpha\beta},\phi\right)+\omega(\phi)\phi_{;\alpha}\phi^{;\alpha}
\right)+\mathcal{L}_{m}\right],
\end{equation}
where $\mathcal{L}_{m}$ and $\omega(\phi)$ are the matter Lagrangian density
and a generic function of the scalar field $\phi$ respectively.

By varying the action (\ref{2.1}) with respect to metric $g_{\mu\nu}$, the
field equations obtained are:
\begin{eqnarray}\label{2.2}
&&\nonumber f_{R}R_{\mu\nu}-\frac{1}{2}\left(f+\omega(\phi) \phi_{;\alpha}
\phi^{;\alpha}\right)g_{\mu\nu}-f_{R;\mu\nu}+g_{\mu\nu}\Box f_{R}+2f_{Y}
R_{\mu}^{\alpha}R_{\alpha\nu}\\
&&-2[f_{Y}R^{\alpha}_{(\mu}]_{;\nu)\alpha}+\Box [f_{Y}R_{\mu\nu}]+[f_{Y}
R_{\alpha\beta}]^{;\alpha\beta}g_{\mu\nu}+\omega(\phi)\phi_{;\mu}\phi_{;\nu}
=\kappa T_{\mu\nu}^{(m)},\label{Fieldeq}
\end{eqnarray}
where $\Box=g^{\mu\nu}\nabla_{\mu}\nabla_{\nu}$, $f_{R}=\partial f/\partial R$
and $f_{Y}=\partial f/\partial Y$. The
energy-momentum tensor for a perfect fluid is defined as
\begin{equation}\label{2.3}
T_{\mu\nu}^{(m)}= (\rho_{m} + p_{m})u_{\mu}u_{\nu} + p_{m}g_{\mu\nu},
\end{equation}
where $p_{m}$, $\rho_{m}$ and $u_{\mu}$ are the pressure, energy density and
the four velocity of the fluid respectively. Hereafter, we will assume that
the matter of the universe has zero pressure $p_{m}=0$ (dust). An effective
Einstein field equation from Eq.~(\ref{2.2}) can be written as
\begin{equation}\label{2.4}
R_{\mu\nu}-\frac{1}{2}Rg_{\mu\nu}=8\pi G_{eff}T_{\mu\nu}^{(m)}+T_{\mu\nu}^{(d)},
\end{equation}
where
\begin{equation}\label{2.5}
G_{eff}=\frac{G}{f_{R}}\,,
\end{equation}
where $G_{eff}$ is the effective
gravitational matter and
\begin{eqnarray}\label{2.6}
\nonumber&&T_{\mu\nu}^{(d)}=\frac{1}{f_{R}}\bigg[-\frac{1}{2}R g_{\mu\nu}f_{R}
+\frac{1}{2}\left(f+\omega(\phi)\phi_{;\alpha}\phi^{;\alpha}\right)g_{\mu\nu}
+f_{R}{}_{;\mu\nu}-g_{\mu\nu}\Box f_{R}-2f_{Y}\\
&&\times R_{\mu}^{\alpha}R_{\alpha\nu}+2[f_{Y}R^{\alpha}_{(\mu}]_{;\nu)\alpha}
-\Box[f_{Y}R_{\mu\nu}]-[f_{Y}R_{\alpha\beta}]^{;\alpha\beta}g_{\mu\nu}-\omega
(\phi)\phi_{;\mu}\phi_{;\nu}\bigg],
\end{eqnarray}
represents an effective energy-momentum tensor related with all the new terms
of the theory. The metric describing the FRW universe is
\begin{equation}\label{2.7}
ds^{2}=h_{\alpha\beta}dx^{\alpha}dx^{\beta}+\tilde{r}^{2}d\Omega^{2},
\end{equation}
with the 2-dimensional metric
$h_{\alpha\beta}=\textrm{diag}\left(-1,\frac{a(t)^{2}}{1-k r^{2}} \right)$,
$(x^{0},x^{1})=(t,r)$, $a(t)$ is the scale factor and $k=\pm1,0$ is the
spacial curvature. The second term is $\tilde{r}=a(t)r$ and
$d\Omega^{2}=d\theta^2+\sin\theta^2d\varphi^2$ is the 2-dimensional sphere
with unit radius. The gravitational field equations for the metric
(\ref{2.7}) are given by
\begin{eqnarray}\label{2.8}
\nonumber3\left(H^{2}+\frac{k}{a^{2}}\right)&=&8\pi G_{eff}\rho_{m}+\frac{1}
{f_{R}}\bigg[\frac{1}{2}\left(Rf_{R}-f\right)-\frac{1}{2}\omega(\phi)
\dot{\phi}^{2}\\
\nonumber&&-3H\partial_{t}f_{R}-6H\big(2\dot{H}+3H^{2}+\frac{k}{a^{2}}\big)
\partial_{t}f_{Y}\\
&&-f_{Y}\bigg(\dddot{H}+4H\ddot{H}+6\dot{H}H^{2}-2H^{4}-\frac{4kH^{2}}
{a^{2}}\bigg)\bigg]\,,\\\label{2.9}
\nonumber-\left(2\dot{H}+3H^{2}+\frac{k}{a^{2}}\right)&=&\frac{1}{f_{R}}\bigg[
\frac{1}{2}\left(f-R f_{R}\right)-\frac{1}{2}\omega(\phi)\dot{\phi}^{2}
+\partial_{tt}f_{R}+2H\partial_{t}f_{R}\\
\nonumber&&+\left(4\dot{H}+6H^{2}+\frac{2k}{a^{2}}\right)\partial_{tt}f_{Y}
+4H\bigg(\dot{H}+3H^{2}\\
\nonumber&&+\frac{2k}{a^{2}}\bigg)\partial_{t}f_{Y}+f_{Y}\bigg(4\dddot{H}
+20H\ddot{H}+10\dot{H}H^{2}\\
&&+16\dot{H}^{2}-18H^{4}-\frac{18k\dot{H}}{a^{2}}-\frac{20kH^{2}}{a^{2}}
-\frac{18k^{2}}{a^{4}}\bigg)\bigg]\,.
\end{eqnarray}
Here, dots and $\partial_{t}$ represent total derivation and partial derivation
with respect to the cosmic time $t$ and $H=\dot{a}/a$ is the Hubble parameter.
These equations can be rewritten as
\begin{eqnarray}\label{2.10}
3\left(H^{2}+\frac{k}{a^{2}}\right)&=&8\pi G_{eff}(\rho_{m}+\rho_{d})\,,\\
\label{2.11}
-2\left(\dot{H}-\frac{k}{a^{2}}\right)&=&8\pi G_{eff}\left(\rho_{m}+\rho_{d}
+p_{d}\right),
\end{eqnarray}
where $\rho_{d}$ and $p_{d}$ are the energy density and pressure of dark
components with $G=f_{R} G_{eff}$, are given by
\begin{eqnarray}\label{2.12}
\nonumber\rho_{d}&=&\frac{1}{8\pi G}\bigg[\frac{1}{2}\left(R f_{R}-f\right)-
\frac{1}{2}\omega(\phi)\dot{\phi}^{2}-3H\partial_{t}f_{R}-6H\bigg(2\dot{H}+3
H^{2}+\frac{k}{a^{2}}\bigg)\\
&&\times\partial_{t}f_{Y}-f_{Y}\bigg(\dddot{H}+4H\ddot{H}+6\dot{H}H^{2}-2H^{4}
-\frac{4kH^{2}}{a^{2}}\bigg)\bigg],
\end{eqnarray}
\begin{eqnarray}\label{2.13}
\nonumber p_{d}&=&\frac{1}{8\pi G}\bigg[\frac{1}{2}\left(f-R f_{R}\right)
-\frac{1}{2}\omega(\phi)\dot{\phi}^{2}+\partial_{tt}f_{R}+2H\partial_{t}f_{R}
+\bigg(4\dot{H}+6H^{2}\\
\nonumber&&+\frac{2k}{a^{2}}\bigg)\partial_{tt}f_{Y}+4H\bigg(\dot{H}+3H^{2}
+\frac{2k}{a^{2}}\bigg)\partial_{t}f_{Y}+f_{Y}\bigg(4\dddot{H}+20H\ddot{H}\\
&&+10\dot{H}H^{2}+16\dot{H}^{2}-18H^{4}-\frac{18k\dot{H}}{a^{2}}-\frac{20k
H^{2}}{a^{2}}-\frac{8k^{2}}{a^{4}}\bigg)\bigg].
\end{eqnarray}
For a dark fluid, the equation of state (EoS) parameter $\omega_{d}$ can be
derived as $(\omega_{d}=\frac{p_{d}}{\rho_{d}})$
\begin{eqnarray}\label{2.14}
\nonumber\omega_{d}&=&-1+\frac{1}{\rho_{d}}\bigg[-\omega(\phi)\dot{\phi}^{2}
+\partial_{tt}\Psi-H\partial_{t}\Psi+\bigg(4\dot{H}+6H^{2}+\frac{2k}{a^{2}}
\bigg)\partial_{tt}f_{Y}\\
\nonumber&&-2H\bigg(4\dot{H}+3H^{2}+\frac{k}{a^{2}}\bigg)\partial_{t}f_{Y}
+f_{Y}\bigg(3\dddot{H}+16H\ddot{H}+4\dot{H}H^{2}+16\dot{H}^{2}\\
&&-16H^{4}-\frac{18k\dot{H}}{a^{2}}-\frac{16kH^{2}}{a^{2}}-\frac{8k^{2}}{a^{4}}
\bigg)\bigg].
\end{eqnarray}
For ordinary matter, the semi-conservation equation is given by
\begin{equation}\label{2.15}
\dot{\rho}+3H\rho=q.
\end{equation}
For dark component, the conservation equation is given by
\begin{eqnarray}\label{2.16}
\dot{\rho}_{d}+3H(\rho_{d}+p_{d})&=&q_{d},\\\label{2.17}
\dot{\rho}_{\rm total}+3H(\rho_{\rm total}+p_{\rm total})&=&q_{\rm total},
\end{eqnarray}
where $\rho_{\rm total}=\rho_{m}+\rho_{d}$, $p_{\rm total}=p_{m}+p_{d}$,
$q_{d}$ is the energy exchange term of dark components and $q_{\rm
total}=q+q_{d}$ is the total energy exchange term. Substituting Eq's.
(\ref{2.10}) and (\ref{2.11}) in the above equation, we have
\begin{equation}\label{2.17}
q_{\rm total}=\frac{3}{8\pi G}\left(H^{2}+\frac{k}{a^{2}}\right)\partial_{t}f_R.
\end{equation}
The energy exchange term for $f(R)$ gravity can be recovered by setting
$f(R,Y,\phi)=f(R)$. In GR, by choosing $f(R,Y,\phi)=R$ we get $q_{\rm total}=0$.

\section{Generalized Thermodynamics laws with non-equilibrium description}\label{sec3}

Here, we discuss the first and second laws of thermodynamics at the apparent
horizon of FRW universe in a more general $f(R, Y, \phi)$ gravity.

\subsection{First Law of Thermodynamics}

In this section, we analyse the validity of the first law of thermodynamics
at the apparent horizon in a FRW universe for $f(R,Y,\phi)$ gravity. The
dynamical apparent horizon is derived by the relation $h^{\alpha\beta}
\partial_{\alpha} \tilde{r}\partial_{\beta}\tilde{r}=0$ from which we have
that the radius of apparent horizon is $\tilde{r}_{A}=\left(H^{2}+k/a^2
\right)^{-\frac{1}{2}}.$ By taking time derivatives in $\tilde{r}_{A}$ and
using Eq. (\ref{2.11}), we obtain
\begin{equation}\label{3.5}
f_{R} d\tilde{r}_{A}=4\pi GH\tilde{r}_{A}^{3}(\hat{\rho}_{\rm total}
+\hat{p}_{\rm total})dt\,,
\end{equation}
where $\hat{\rho}_{\rm total}=\hat{\rho_{m}}+\hat{\rho_{d}}$ and
$\hat{p}_{\rm total}=\hat{p_{d}}$ are the total energy density and total
pressure respectively. Here, $d\tilde{r}_{A}$ represents the infinitesimal
change in the radius of the apparent horizon during an infinitesimal time
interval $dt$. The temperature of the apparent horizon is defined as
$T_{h}=|K_{sg}|/(2\pi)$, where $K_{sg}$ is the surface gravity given by
\cite{24}
\begin{align}
K_{sg}=-\frac{1}{\tilde{r}_{A}}\Big(1-\frac{\dot{\tilde{r}}_{A}}{2H
\tilde{r}_{A}}\Big)\,.
\end{align}
In GR, Bekenstein and Hawking  defined the horizon entropy by the relation
$S_{h}=A/4G$, where $A$ is the area of the apparent horizon defined by
$A=4\pi\tilde{r}_{A}^{2}$  \cite{9,14,17}. In the literature of modified
theories of gravity, Wald introduced the horizon entropy with a Noether
charge \cite{38}. This quantity  can be obtained by varying the Lagrangian
density of the modified theory with respect to Riemann tensor. This entropy
is defined as $\hat{S_{h}}=A/4G_{eff}$ \cite{39}, where $G_{eff}$ is the
effective gravitational coupling. The Wald's entropy in $f(R,Y,\phi)$ gravity
is defined as
\begin{equation}\label{3.4}
\hat{S_{h}}=\frac{f_{R}A}{4G}\,.
\end{equation}
By differentiating Eq. (\ref{3.4}) and using (\ref{3.5}), we
have
\begin{equation}\label{3.6}
\frac{1}{2\pi \tilde{r}_{A}}d\hat{S_{h}}=4\pi\tilde{r}_{A}^{3}
\left(\hat{\rho}_{\rm total}+\hat{p}_{\rm total}\right)Hdt
+\frac{\tilde{r}_{A}}{2 G}df_{R}.
\end{equation}
If we multiply both sides of the above equation by
$1-\dot{\tilde{r}}_{A}/(2 H \tilde{r}_{A})$, we obtain
\begin{equation}\label{3.7}
T_{h} d\hat{S_{h}}=-4 \pi\tilde{r}_{A}^{3}\left(\hat{\rho}_{\rm total}
+\hat{p}_{\rm total}\right)H dt+2\pi\tilde{r}_{A}^{2}\left(\hat{\rho}_{\rm
total}+\hat{p}_{\rm total}\right)d\tilde{r}_{A}+\frac{\pi\tilde{r}_{A}^{2}
T_{h}}{G} df_{R}.
\end{equation}
Now, we will define the energy of the universe inside the apparent horizon.
The Misner-Sharp energy defined in \cite{40} is $E=\tilde{r}_{A}/(2G)$ and
for $f(R,Y,\phi)$ gravity we can write it as \cite{42}
\begin{equation}\label{3.8}
\hat{E}=\frac{\tilde{r}_{A}}{2G_{eff}}\,,
\end{equation}
We can also rewrite this expression by using the volume $V=(4/3)\pi
\tilde{r}_{A}^{3}$, yielding
\begin{equation}\label{3.9}
\hat{E}=\frac{3V}{8\pi G_{eff}}\left(H^{2}+\frac{k}{a^{2}}\right)=V
\hat{\rho}_{\rm total}\,,
\end{equation}
which is the total energy inside the sphere of radius $\tilde{r}_{A}$. If we
choose the effective gravitational coupling constant as positive in
$f(R,Y,\phi)$ gravity then we have $G_{eff}=G/f_{R}>0$, from which we can
conclude that $\hat{E}>0$. From Eqs. (\ref{2.10}) and (\ref{3.9}) we will
have that
\begin{equation}\label{3.10}
d\hat{E}=4\pi\tilde{r}_{A}^{2}\hat{\rho}_{\rm total}d\tilde{r}_{A}-4\pi
\tilde{r}_{A}^{3}\left(\hat{\rho}_{\rm total}+\hat{p}_{\rm total}\right)Hdt
+\frac{\tilde{r}_{A}}{2G}df_{R}\,.
\end{equation}
Using Eq. (\ref{3.10}) in (\ref{3.7}), it follows that
\begin{equation}\label{3.11}
T_{h}d\hat{S}_{h}=d\hat{E}-\hat{W}dV-\frac{\left(1-2\pi\tilde{r}_{A}T_{h}
\right)\tilde{r}_{A}}{2G}df_{R}\,,
\end{equation}
where we have used the work density
$\hat{W}=(1/2)(\hat{\rho}_{\rm total}-\hat{p}_{\rm total})$ \cite{43}.
The above equation can be rewritten as
\begin{equation}\label{3.12}
T_{h}d \hat{S}_{h}+T_{h}d_{i}\hat{S}_{h}=d \hat{E}-\hat{W} dV\,,
\end{equation}
where
\begin{equation}\label{3.13}
d_{i}\hat{S}_{h}=\frac{\left(1-2\pi\tilde{r}_{A}T_{h}\right)\tilde{r}_{A}}{2G
T_{h}}df_{R}=\frac{(\hat{E}-\hat{S}_{h}T_{h})}{T_{h}}\frac{df_{R}}{f_{R}}\,.
\end{equation}
If we compare the expression above for $f(R,Y,\phi)$ gravity with GR,
Lovelock gravity and Gauss-Bonnet gravity, we see an additional term
$d_{i}\hat{S}_{h}$ in the first law of thermodynamics. We can call that extra
term as the entropy production term which occurs due to the non-equilibrium
behaviour of $f(R,Y,\phi)$ gravity. From this result, by setting
$f(R,Y,\phi)=f(R)$ we recover the first law of thermodynamics in
non-equilibrium of $f(R)$ gravity \cite{29}. Moreover, if we choose
$f(R,Y,\phi)=R$, we can achieve the standard first law of thermodynamics in
GR.

\subsection{Generalized Second Law of Thermodynamics}

In modified gravitational theories, the Generalized Second law of
Thermodynamics (GSLT) has been widely discussed
\cite{29}-\cite{32,42,48,Zubair:2016uhx}. In order to check its validity in
$f(R,Y,\phi)$ gravity, we have to prove the inequality \cite{42}
\begin{equation}\label{3.14}
\dot{\hat{S_{h}}}+d_{i}\dot{\hat{{S}_{h}}}+\dot{\hat{S}}_{\nu}\geq0\,,
\end{equation}
where $\hat{S_{h}}$, $d_{i}\dot{\hat{{S}_{h}}}=\partial_{t}(d_{i}
\hat{S_{h}})$ and $\hat{S}_{\nu}$ are the horizon entropy, entropy due to all
the matter inside the horizon and the entropy due to energy sources inside
the horizon respectively. The Gibb's equation which includes the entropy of
matter and energy fluid is given by \cite{50}
\begin{equation}\label{3.15}
T_{\nu}d\hat{S_{\nu}}=d(\hat{\rho}_{\rm total}V)+\hat{p}_{\rm total}dV\,,
\end{equation}
where $T_{\nu}$ denotes the temperature within the horizon. Now, we will
assume a relation between the temperature within the horizon and the
temperature of the apparent horizon given by
\begin{equation}
T_{\nu}=bT_{h}\,,
\end{equation}
where $b$ is a constant which lies between $0<b<1$ to guarantee the
positivity of the temperature and also to have a smaller temperature than the
horizon one. By substituting Eqs. (\ref{3.12}) and (\ref{3.15}) in Eq.
(\ref{3.14}), we obtain
\begin{equation}\label{3.16}
\dot{S}_{tot}=\dot{\hat{S_{h}}}+d_{i}\dot{\hat{{S}_{h}}}+\dot{\hat{S_{\nu}}}=
\frac{2\pi\Sigma}{\tilde{r}_{A}bR}\geq0\,,
\end{equation}
where
\begin{equation}
\nonumber\Sigma=(1-b)\dot{\hat{\rho}}_{\rm total}V+(1-\frac{b}{2})
(\hat{\rho}_{\rm total}+\hat{p}_{\rm total})\dot{V}\,,
\end{equation}
which is the general condition to satisfy the GSLT in modified gravitational
theories \cite{42}. Using Eqs. (\ref{2.10}) and (\ref{2.11}), the condition
(\ref{3.16}) is reduced to
\begin{equation}\label{3.17}
\frac{2\pi\Xi}{Gb\left(H^{2}+\frac{k}{a^{2}}\right)\left(\dot{H}+2H^{2}
+\frac{k}{a^{2}}\right)}\geq0,
\end{equation}
where
\begin{eqnarray}
\nonumber\Xi&=&(b-1)\partial_{t}f_{R}\left(H^{2}+\frac{k}{a^{2}}\right)+2Hf_{R}
(b-1)\left(\dot{H}-\frac{k}{a^{2}}\right)+(b-2)\\
&&\times f_{R}H\left(\dot{H}-\frac{k}{a^{2}}\right)^{2}\left(H^{2}+\frac{k}
{a^{2}}\right)^{-1}.
\end{eqnarray}
In case of flat FRW universe, the GSLT is satisfied with the conditions
$\partial_{t}f_{R}\geq0$, $f_{R}\geq0$, $H\geq0$ and $\dot{H}\geq0$. To
protect the GSLT, the condition (\ref{3.17}) is equivalent to $\Xi\geq0$.

\section{Validity of GSLT} \label{sec4}

Here, we will employ some interesting models in $f(R, Y, \phi)$ gravity
(reconstructed in \cite{34a}) and also some specific forms of $f(R, \phi)$ in
order to check the validity of the GSLT, $\dot{S}_{tot}\geq0$ for different
cosmological solutions.

\subsection{Model constructed from de-Sitter Universe}

To explain the current cosmic era in cosmology, the de-Sitter solution is
very important. In de-Sitter universe, the scale factor, Hubble parameter,
Ricci tensor and the scalar field are defined as \cite{13a}
\begin{equation}\label{1.2}
a(t)=a_{0}e^{H_{0}t},~~ H=H_{0},~~R=12H_{0}^{2}~~ \textrm{and}~~ \phi(t)\sim
a(t)^{\beta}\,.
\end{equation}
\begin{itemize}
\item \textbf{de-Sitter model $f(R,Y,\phi)$}
\end{itemize}
We have constructed the more general model $f(R,Y,\phi)$ in \cite{34a}, here
we are using this model to examine the viability of the GSLT. For this model,
the function reads
\begin{equation*}
f(R,Y,\phi)=\alpha_{1}\alpha_{2}\alpha_{3}e^{\alpha_{1}R}e^{\alpha_{2}Y}
\phi^{\gamma_{1}}+\gamma_{2}\phi^{\gamma_{3}}+\gamma_{4}\phi^{\gamma_{5}}\,,
\end{equation*}
where $\alpha_{i}$ are integration constants and we have defined
\begin{eqnarray*}
\gamma_{1}&=&\frac{18\beta\alpha _1 H_0^{2}-108\beta\alpha _2 H_{0}^4-5+6
\alpha _1H_{0}^2-84\alpha _2H_{0}^4}{6\left(H_{0}^2\alpha _1\beta-6\beta
\alpha_2H_{0}^4\right)}\\
\gamma_{2}&=&\omega_{0}\beta^{2} H_{0}^{2},~~\gamma_{3}=m+2,~~\gamma_{4}=-2
\kappa\rho_{0} a_{0}^{3},~~\gamma_{5}=-\frac{3}{\beta}.
\end{eqnarray*}
Introducing this model in (\ref{3.16}), the validity of the GSLT will hold if
\begin{eqnarray}
&&\nonumber\dot{S}_{tot}=\frac{2\pi}{Gb}\left[-12kH_{0}(b-1)\alpha_{1}^{3}
\alpha_{2}\alpha_{3}a_{0}^{\beta\gamma_{1}}e^{\alpha_{1}R+\alpha_{2}Y}\left
(a_{0}^2H_{0}^2+ke^{-2H_{0}t}\right)e^{-2H_{0}t}\right.\\
&&\nonumber\left.\times e^{\beta\gamma_{1}H_{0}t}-\frac{24}{a_{0}^2}k H_{0}
(b-1)\left(a_{0}^2H_{0}^2+ke^{-2H_{0}t}\right)\left(3a_{0}H_{0}^2e^{-3H_{0}t}
+2k e^{-4H_{0}t}\right)\alpha_{1}^{2}\right.\\
&&\nonumber\left.\times\alpha_{2}^{2}\alpha_{3}a_{0}^{\beta\gamma_{1}}
e^{\alpha_{1}R+\alpha_{2}Y}e^{\beta\gamma_{1}H_{0}t}+\beta H_{0}(b-1)
\alpha_{1}^{2}\alpha_{2}\alpha_{3}\gamma_{1}a_{0}^{\beta+2}e^{\beta H_{0}t}
\left(a_{0}^2H_{0}^2\right.\right.\\
&&\nonumber\left.\left.+ke^{-2H_{0}t}\right)e^{\alpha_{1}R+\alpha_{2}Y}
a_{0}^{\beta(\gamma_{1}-1)}e^{\beta(\gamma_{1}-1)H_{0}t}-2kH_{0}a_{0}^2(b-1)
\alpha_{1}^{2}\alpha_{2}\alpha_{3}e^{-2H_{0}t}a_{0}^{\beta
\gamma_{1}}\right.\\
&&\nonumber\left. \times e^{\alpha_{1}R+\alpha_{2}Y} e^{\beta\gamma_{1}H_{0}t}
+2k^2 H_{0}a_{0}^2\left(\frac{b}{2}-1\right)\alpha_{1}^{2}\alpha_{2}\alpha_{3}
\left(a_{0}^2H_{0}^2+ke^{-2H_{0}t}\right)^{-1}e^{-4H_{0}t}\right.\\
&&\left.\times e^{\alpha_{1}R+\alpha_{2}Y} a_{0}^{\beta\gamma_{1}}e^{\beta
\gamma_{1} H_{0}t}\right]\left(a_{0}^2H_{0}^2+ke^{-2H_{0}t}\right)^{-1}
\left(2a_{0}^2 H_{0}^2+ke^{-2H_{0}t}\right)^{-1}\geq0\,.
\end{eqnarray}
The validity of the GSLT in de-Sitter $f(R,Y,\phi)$ depends on five
parameters $\alpha_{1}$, $\alpha_{2}$, $\alpha_{3}$, $\beta$ and $t$. In this
perspective, we can fix two parameters and observe the feasible region by
varying the possible ranges for the other parameters. In our case, we will
fix the parameters $\alpha_{1}$ and $\alpha_{2}$ and show the results for
$\dot{S}_{tot}$. Herein, we set the present day values of the Hubble
parameter and the cosmographic parameters as $H_{0}=67.3$, $q=-0.81$,
$j=2.16$, $s=-0.22$ \cite{32a}. The feasible regions for all the possible
cases for de-Sitter $f(R,Y,\phi)$ model are presented in Table \ref{Table1}.

\begin{figure}[H]
\centering \epsfig{file=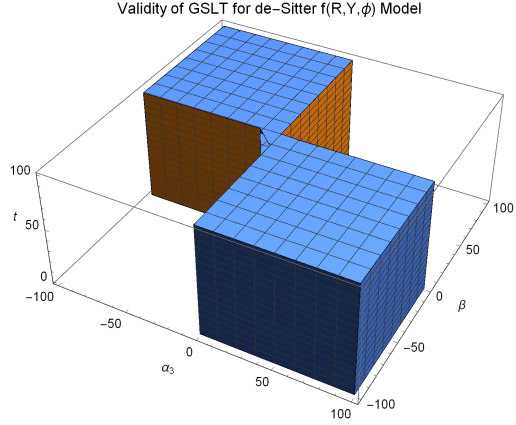,
width=.49\linewidth,height=2.2in} \caption{Validity region of the
GSLT for de-Sitter $f(R,Y,\phi)$ model with $\alpha_1=1$ and
$\alpha_2=3$.} \label{Fig1}
\end{figure}

Initially, we will vary $\alpha_{1}$ and $\alpha_{2}$ to check the validity
of $\dot{S}_{tot}$ for different values of $\alpha_{3}$, $\beta$ and $t$. If
we set both $\alpha_{1}$ and $\alpha_{2}$ as positive then
$\dot{S}_{tot}\geq0$ is valid at every time, however $\alpha_{3}$ and $\beta$
must be in the ranges ($\alpha_{3}\geq0$, $\beta\leq-0.78$) and
($\alpha_{3}\leq0$, $\beta\geq0$). If $\alpha_{1}<0$ and $\alpha_{2}>0$, the
validity of the GSLT holds at all times with ($\alpha_{3}\geq0$,
$\beta\leq-0.78$) or ($\alpha_{3}\leq0$, $\beta\geq0$). For ($\alpha_{1}>0$,
$\alpha_{2}<0$), $\dot{S}_{tot}\geq0$ is valid for all values of
$\alpha_{3}$, $\beta$ and $t$. For ($\alpha_{1}>0$, $\alpha_{2}<0$) and
($\alpha_{1}<0$, $\alpha_{2}<0$), the validity of the GSLT is true for all
values of $\alpha_{3}$, $\beta$ and $t$. As an example, Fig.~\ref{Fig1} shows
the evolution of the GSLT constraint with the parameters $\alpha_3$, $\beta$
and $t$ by fixing $\alpha_1>0$ and $\alpha_2>0$.

\begin{itemize}
\item de-Sitter model independent of Y
\end{itemize}
Now we are considering a function $f(R,\phi)$ independent of $Y$ and the
model constructed in \cite{34a} is defined as
\begin{equation}\label{1.1}
f(R,\phi)=\alpha_{1}\alpha_{2}e^{\alpha_{1}R}\phi^{\gamma_{1}}+\gamma_{2}
\phi^{\gamma_{3}}+\gamma_{4}\phi^{\gamma_{5}}\,,
\end{equation}
where $\alpha_{i}'s$ are constants of integration and
\begin{eqnarray}\nonumber
\nonumber\gamma_{1}&=&-\frac{1}{\beta}\Big(1+\frac{1}{6H_{0}^{2}\alpha_{1}}
\Big),~~\gamma_{2}=\omega_{0}\beta^{2} H_{0}^{2}\,,\\ \nonumber
\gamma_{3}&=&m+2,~~\gamma_{4}=-2 \kappa^2 \rho_{0}a_{0}^{3(1+w)},~~\gamma_{5}=
-\frac{3}{\beta}\,.
\end{eqnarray}
Inserting this model in (\ref{3.16}), the viability of the GSLT is given by
\begin{eqnarray}
&&\nonumber\dot{S}_{tot}=\frac{2\pi}{Gb}\bigg[-12kH_{0}(b-1)\alpha_{1}^{3}
\alpha_{2}a_{0}^{\beta\gamma_{1}}e^{\alpha_{1}R}e^{-2H_{0}t}e^{\beta\gamma_{1}
H_{0}t}+\beta H_{0}(b-1)\alpha_{1}^{2}\alpha_{2}\\
&&\nonumber\times\gamma_{1}\left(a_{0}^2H_{0}^2+ke^{-2H_{0}t}\right)a_{0}^2
a_{0}^{\beta\gamma_{1}}e^{\alpha_{1}R}e^{\beta\gamma_{1}H_{0}t}-2kH_{0}(b-1)
\alpha_{1}^{2}\alpha_{2}a_{0}^2 a_{0}^{\beta\gamma_{1}}e^{\alpha_{1}R}\\
&&\nonumber\times e^{\beta\gamma_{1}H_{0}t}e^{-2H_{0}t}+2k^2H_{0}a_{0}^2
e^{-4H_{0}t}\left(\frac{b}{2}-1\right)\left(H_{0}^2a_{0}^2+ke^{-2H_{0}t}
\right)^{-1}\alpha_{1}^2\alpha_2 e^{\alpha_{1}R}\\
&&\times e^{\beta\gamma_{1}H_{0}t}a_{0}^{\beta\gamma_{1}}\bigg]\left(a_{0}^2
H_{0}^2+ke^{-2H_{0}t}\right)^{-1}\left(2a_{0}^2H_{0}^2+ke^{-2H_{0}t}
\right)^{-1}\geq0\,.
\end{eqnarray}
The viability of the GSLT in de-Sitter $f(R,\phi)$ depends on four
parameters $\alpha_{1}$, $\alpha_{2}$, $\beta$ and $t$. By fixing $\beta$ we
will observe the variations of $\alpha_{1}$ and $\alpha_{2}$ where GSLT is
valid. For all values of $\beta$ and $t$ we have two cases $(i)$
$\alpha_{1}\leq-0.1$ $\forall$ $\alpha_{2}$ $(ii)$ $\alpha_{1}>0$ with
$\alpha_{2}\geq0$.

\subsection{Model constructed from power Law method}

Power solutions are very useful to discuss the different phases of cosmic
evolution e.g., dark energy, matter and radiation dominated epochs. We are
discussing here just one power law solution for $f(R,Y,\phi)$ gravity with
a power law scale factor defined as \cite{32a,33a}
\begin{equation}\label{25}
a(t)=a_{0}t^{n},~~H(t)=\frac{n}{t},~~R = 6n(1 - 2n)t^{-2}\,,
\end{equation}
where $n>1$ shows the accelerating picture of the universe, $0<n<1$ leads to
decelerated universe, $n=2/3$ leads to dust dominated era and $n=1/2$ for
radiation dominated epoch.
\begin{itemize}
\item \textbf{Power Law model independent of Y}
\end{itemize}
In \cite{34a}, it was constructed a well-behaved model for $f(R,\phi)$
function, so that now we will concentrate in that specific theory and model
to show the validity of the GSLT. For this model, the $f(R,\phi)$ function
reads
\begin{equation*}
f(R,\phi)=\alpha_{1}\alpha_{2}\phi^{\gamma_{1}}R^{\gamma_{2}}+\gamma_{3}
\phi^{\gamma_{4}}+\gamma_{5}\phi^{\gamma_{6}}\,,
\end{equation*}
where $\alpha_{i}$ are integration constants and
\begin{eqnarray*}
\gamma_{1}&=&\frac{\alpha_{1}}{3n-1}+\frac{n-3}{n\beta}-\frac{2(3n-1)^{2}}
{n^{2}\beta^{2}\alpha_{1}},~~~\gamma_{2}=\frac{n(n-3)\beta\alpha_{1}}
{(3n-1)^{2}},\\
\gamma_{3}&=&\omega_{0}\beta^{2}n^{2}a_{0}^{\frac{2}{n}},~~~\gamma_{4} = m+2
-\frac{2}{n\beta},~~~\gamma_{5}=-2 \kappa \rho_{0}a_{0}^{3(1+w)},
~~~\gamma_{6}=-\frac{3}{\beta}.
\end{eqnarray*}
By substituting this model in (\ref{3.16}) we get
\begin{eqnarray}
&&\nonumber\dot{S}_{tot}=\frac{2\pi}{Gb}\bigg[H_{0}(b-1)\alpha_{1}\alpha_{2}
\gamma_{2}(\gamma_{2}-1)a_{0}^{\beta\gamma_{1}}t^{n\beta\gamma_{1}}\left(\left
(j-q-2\right)H_{0}^2+\frac{k}{a_{0}^2t^{2n}}\right)\\
&&\nonumber\times\left((1-q)H_{0}^2+\frac{k}{a_{0}^2t^{2n}}\right)^{\gamma_{2}
-2}+\beta H_{0}\alpha_{1}\alpha_{2}\gamma_{1}\gamma_{2}\left((1-q)H_{0}^2+
\frac{k}{a_{0}^2 t^{2n}}\right)^{\gamma_{2}-1}\\
&&\nonumber\times(b-1)a_{0}^{\beta\gamma_{1}}t^{n\beta\gamma_{1}}-2H_{0}(b-1)
\alpha_{1}\alpha_{2}\gamma_{2}a_{0}^{\beta\gamma_{1}}t^{n\beta\gamma_{1}}\left
((1-q)H_{0}^2+\frac{k}{a_{0}^2 t^{2n}}\right)^{\gamma_{2}-1}\\
&&\nonumber\times\left\{1+\frac{q H_{0}^{2}}{H_{0}^2+\frac{k}{a_{0}^2t^{2n}}}
\right\}+2H_{0}\alpha_{1}\alpha_{2}\gamma_{2}a_{0}^{\beta\gamma_{1}}t^{n\beta
\gamma_{1}}\left((1-q)H_{0}^2+\frac{k}{a_{0}^2 t^{2n}}\right)^{\gamma_{2}-1}\\
&&\times\left(\frac{b}{2}-1\right)\left(1+\frac{qH_{0}^2}{H_{0}^2+\frac{k}
{a_{0}^2t^{2n}}}\right)^2\bigg]\left((1-q)H_{0}^2+\frac{k}{a_{0}^2t^{2n}}
\right)^{-1}\geq0.
\end{eqnarray}
The above constraint has five parameters $\alpha_{1}$, $\alpha_{2}$, $n$,
$\beta$ and $t$. Now, we will check the validity of $\dot{S}_{tot}\geq 0$ for
different values of $n$, $\beta$ and $t$ by fixing $\alpha_{1}$,
$\alpha_{2}$. All possible cases of this model are written in Table
\ref{Table1}.

\begin{figure}[H]
\centering \epsfig{file=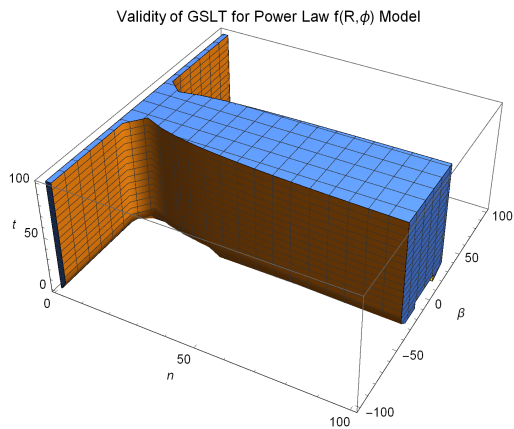, width=.49\linewidth,height=2.2in}
\caption{Valdity of the GSLT for Power law-$f(R,\phi)$ versus the
parameters $n$, $\beta$ and $t$ with $\alpha_{1}=1$ and
$\alpha_{2}=5$.} \label{Fig2}
\end{figure}

Let us start with the case $\alpha_{1}>0$ and check the viable ranges of
$\alpha_{2}$, $\beta$ and $t$. In this case we have three cases depending on
the choice $\alpha_{2}$,\\\\
 $(i)$ $\alpha_{2}<0$, $n\geq3$ with ($\beta\leq-35.8$,
$t\geq1$) or ($\beta\geq2.81$, $t\geq0.94$).\\
 $(ii)$ $\alpha_{2}=0$, $n>1$
and $\forall$ $t$ with $\beta>0$ or $\beta<0$.\\
 $(iii)$ $\alpha_{2}>0$ with
($n\geq8.6$, $0<\beta\leq20$, $t\geq0.8$) or ($n\geq12.7$, $-20\leq\beta<0$,
$t\geq0.9$).\\ \\
If we take $\alpha_{1}<0$ we also have three possible cases where the GSLT
will hold,\\ \\
$(i)$ $\alpha_{2}<0$ with ($n\geq8.6$,
$-20\leq\beta<0$, $t\geq0.8$) or ($n\geq12.7$, $0<\beta\leq20$, $t\geq0.9$).\\
$(ii)$ $\alpha_{2}=0$, $n>1$ and $\forall$ $t$ with $\beta>0$ or $\beta<0$.\\
$(iii)$ $\alpha_{2}>0$, $n\geq3$ and $\forall$ $t$ with ($\beta\leq-28.1$)
or ($\beta\geq35.7$). \\ \\
It can be noted that by taking $\alpha_{1}$ and $\alpha_{2}$ with the same
sign, $\dot{S}_{tot}\geq0$ is not valid for initial values of $t$ and $n$,
and also that $\beta$ is restricted to $\beta\leq20$ or $\beta\geq-20$. Fig.
(\ref{Fig2}) depicts the validity region of the GSLT for a specific case in
this model, where $\alpha_{1}>0$ and $\alpha_{2}=5$ .

\subsection{$f(R,\phi)$ Models}
In this section, we will study the validity of the GSLT using well known
forms of $f(R,\phi)$ gravity. 
\subsubsection{Model-I}

Myrzakulov et al. \cite{35a} have examined the spectral index and
tensor-to-scalar ratio to describe the inflation in $f(R,\phi)$ theories and
observed the results using the recent observational data. This model is based
on the following function,
\begin{equation}
\nonumber f(R,\phi)=\frac{R-2\Lambda(1-e^{B\phi\kappa^{3}R})}{\kappa^{2}}\,,
\end{equation}
where $\kappa^{3}$ is introduced for dimensional reasons and $B$ is a
constant. Inserting the model in (\ref{3.16}), the inequality becomes
\begin{eqnarray}
&&\nonumber\dot{S}_{tot}=\frac{2\pi}{Gb}\bigg[2H_{0}(b-1)\Lambda B^2 a_{0}^{2
\beta}t^{2n\beta}\kappa^4Re^{Ba_{0}^\beta t^{n\beta}\kappa^3 R}+2\Lambda B
\kappa(b-1)\beta H_{0}a_{0}^{\beta} t^{n\beta}\\
&&\nonumber\times\left(1+Ba_{0}^\beta t^{n\beta}\kappa^{3}R\right)
e^{B a_{0}^\beta t^{n\beta}\kappa^3R}-\frac{2}{\kappa^2}H_{0}(b-1)
\left\{1+2\Lambda B a_{0}^\beta t^{n\beta}\kappa^3e^{B a_{0}^\beta t^{n\beta}
\kappa^3R}\right\}\\
&&\nonumber\times\left(1+\frac{q H_{0}^2}{H_{0}^2+\frac{k}{a_{0}^2 t^{2n}}}
\right)+\frac{2H}{\kappa^2}\left(\frac{b}{2}-1\right)\left\{1+2\Lambda B
a_{0}^\beta t^{n\beta}\kappa^3 e^{Ba_{0}^\beta t^{n\beta}\kappa^3 R}\right\}\\
&&\times\left(1+\frac{q H_{0}^2}{H_{0}^2 +\frac{k}{a_{0}^2 t^{2n}}}\right)^2
\bigg]\left(H_{0}^2(1-q)+\frac{k}{a_{0}^2t^{2n}}\right)^{-1}\geq 0\,,
\end{eqnarray}
where $R=6[(1-q)H_{0}^2 +k/(a_{0}^2 t^{2n})]$. As we can notice, the
inequality depends on four parameters $B$, $n$, $\beta$ and $t$. We can see
that $\dot{S}_{tot}\geq 0$ is satisfied for two cases depending on the choice
of $B$: \\ \\
$(i)$ $B=0$ with $n>1$ and $t\geq0.96$
(for all values of $\beta$).\\
$(ii)$ $B>0$ with $n>1$, $\beta\leq-0.6$ and $\forall$
$t$ and $n\geq2.5$, $\beta\geq6$ and $t\geq2.5$.
\subsubsection{Model-II}

\begin{figure}[H]
\centering \epsfig{file=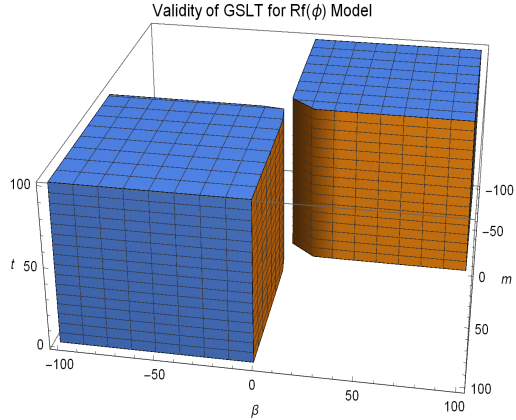,
width=.49\linewidth,height=2.2in} \caption{Regions where the GSLT is
satisfied for the Model-II versus the parameters $m$, $\beta$ and
$t$ with $n=1.1$.}\label{Fig3}
\end{figure}

Now, we will explore a model studied in $\cite{32a}$, where they considered a
function form $f(R,\phi)=Rf(\phi)$ with $\omega(\phi)=\omega_{0}\phi^m$ and
$\phi=a(t)^\beta$. Therefore, we will consider the following function
\begin{equation}
\nonumber f(R,\phi)=R\left(\frac{\omega_{0}\beta^{2}n^{2}a_{0}^{2/n}(mn\beta
+2n\beta+6n-2)}{mn\beta+2n\beta-2} \right)\phi^{m+2-\frac{2}{n\beta}}\,,
\end{equation}
where $\omega_{0}$ and $a_{0}$ are constants. Introducing this model in
(\ref{3.16}) we find the constraint
\begin{eqnarray}
&&\nonumber\dot{S}_{tot}=\frac{2\pi}{Gb}\bigg[\beta^2 H_{0}(b-1)\omega_{0}n
\left(mn\beta+2n\beta+6n-2\right)a_{0}^{\beta(m+2)}t^{mn\beta+2n\beta-2}\\
&&\nonumber-2H_{0}(b-1)\left(\frac{\omega_{0}\beta^2 n^2 a_{0}^{2/n}\left(m n
\beta+2n\beta+6n-2\right)}{mn\beta+2n\beta-2}\right)a_{0}^{\beta\left(m+2
-\frac{2}{n\beta}\right)} t^{mn\beta+2n\beta-2}\\
&&\nonumber\times\left\{1+q H_{0}^2 (H_{0}^2+\frac{k}{a_{0}^2}t^{2n})^{-1}
\right\}+2\left(\frac{b}{2}-1\right)H_{0}a_{0}^{\beta\left(m+2-\frac{2}{n\beta}
\right)} t^{mn\beta+2n\beta-2}\\
&&\nonumber\times\left(\frac{\omega_{0}\beta^2 n^2 a_{0}^{2/n}\left(mn\beta
+2n\beta+6n-2\right)}{mn\beta+2n\beta-2}\right)\left(1+\frac{q H_{0}^2}
{H_{0}^2 +\frac{k}{a_{0}^2 t^{2n}}}\right)^2 \bigg]\\
&&\times\left(H_{0}^2(1-q)+\frac{k}{a_{0}^2t^{2n}}\right)^{-1}\geq 0.
\end{eqnarray}
One can see that the inequality of this model is depending on four parameters
$\beta$, $m$, $n$ and $t$ and hence we will discuss its viability for
different values of $\beta$ and $m$ by fixing $n$. For $n>1$ at every time
$t$ we have two cases where the validity of the GSLT holds: $m\geq2$ with
$\beta\leq-1.5$ and $m\leq-3.2$ with $\beta\geq5$.  In Fig.~\ref{Fig3}, the
validity of the GSLT region is showed for the parameters $m$, $\beta$ and $t$
by fixing $n>1$.
\subsubsection{Model-III}
Now, we will study the model which is applied to describe the cosmological
perturbations for non-minimally coupled scalar field dark energy in both
metric and Palatini formalisms  \cite{36a}.
\begin{equation}
\nonumber f(R,\phi)=R(1+\xi\kappa^{2}\phi^{2}),
\end{equation}
where $\xi$ is the coupling constant. Using this model in (\ref{3.16}) we
have
\begin{eqnarray}
&&\nonumber\dot{S}_{tot}=\frac{2\pi}{Gb}\bigg[2\beta H_{0}\xi \kappa^2(b-1)
a_{0}^{2\beta}t^{2n\beta}-2H_{0}(b-1)\left(1+\xi\kappa^2a_{0}^{2\beta}t^{2n
\beta}\right)\times\\
&&\nonumber\left\{1+q H_{0}^2\left(H_{0}^2+\frac{k}{a_{0}^2}t^{2n}\right)^{-1}
\right\}+2\left(\frac{b}{2}-1\right)H_{0}\left(1+\xi\kappa^2a_{0}^{2\beta}
t^{2n\beta}\right)\times\\
&&\left(1+\frac{q H_{0}^2}{H_{0}^2 +\frac{k}{a_{0}^2 t^{2n}}}\right)^2\bigg]
\times\left(H_{0}^2(1-q)+\frac{k}{a_{0}^2t^{2n}}\right)^{-1}\geq 0.
\end{eqnarray}
Here, we have four parameters $n$, $\xi$, $\beta$ and $t$. Therefore, we can
fix $n$ to find the values of $\xi$ and $\beta$ where the GSLT is satisfied. For
$n>1$, it is valid for $\beta\leq-3.5$ with (for all values of $\xi$ and
$t\geq4$) and for $\beta\geq0.15$ with ($\xi\leq0$ and $t\geq1$).

\subsubsection{Model-IV}

\begin{figure}[H]
\centering\epsfig{file=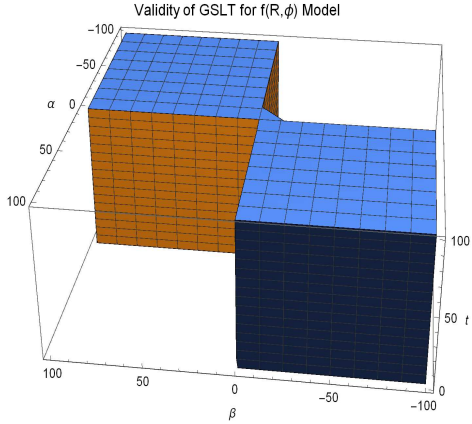, width=.45\linewidth,
height=2.2in}\caption{Validity regions of the  GSLT for the Model-IV
for the parameters $\alpha$ and $\beta$ with $n=1.1$.} \label{Fig4}
\end{figure}

In \cite{37a}, authors used the following model to discuss the inflationary
paradigm
\begin{equation}
\nonumber f(R,\phi)=\phi(R+\alpha R^{2})\,,
\end{equation}
where $\alpha$ is a constant with suitable dimensions. Here, we are
interested to explore the validity of GSLT for the above model. Introducing
this model in (\ref{3.16}) we have inequality of the form
\begin{eqnarray}
&&\nonumber\dot{S}_{tot}=\frac{2\pi}{Gb}\bigg[12\alpha H_{0}(b-1)a_{0}^{\beta}
t^{n\beta}\left\{(j-q-2)H_{0}^2+\frac{k}{a_{0}^2 t^{2n}}\right\}+\beta H_{0}
(b-1)\\
&&\nonumber\times a_{0}^{\beta}t^{n\beta}\left\{1+12\alpha\left((1-q)H_{0}^2
+\frac{k}{a_{0}^2 t^{2n}}\right)\right\}-2H_{0}(b-1)a_{0}^{\beta}t^{n\beta}
\bigg\{1+12\alpha\\
&&\nonumber\times\left((1-q)H_{0}^2+\frac{k}{a_{0}^2 t^{2n}}\right)\bigg\}
\left\{1+q H_{0}^2 \left(H_{0}^2+\frac{k}{a_{0}^2 t^{2n}}\right)^{-1}\right\}
+2\left(\frac{b}{2}-1\right)\\
&&\nonumber \times H_{0}a_{0}^{\beta}t^{n\beta}\left\{1+12\alpha\left((1-q)
H_{0}^2+\frac{k}{a_{0}^2 t^{2n}}\right)\right\}\left(1+\frac{q H_{0}^2}
{H_{0}^2+\frac{k}{a_{0}^2 t^{2n}}}\right)^2 \bigg]\\
&&\times\left(H_{0}^2(1-q)+\frac{k}{a_{0}^2t^{2n}}\right)^{-1}\geq 0.
\end{eqnarray}
We have four parameters $n$, $\alpha$, $\beta$ and $t$ to constraint to satisfy
the above relation. For $n>1$ we have two cases depending on the choice of
$\alpha$:\\ \\
$(i)$ $\alpha<0$ with $\beta\geq0$ (for all times $t$).\\
$(ii)$ $\alpha\geq0$ with $\beta\leq-0.25$ and $t\geq1$.\\ \\
In (\ref{Fig4}) is showed the evolution of the GSLT in this model for a
specific choice of the parameter $n=1.1$

\section{Equilibrium description of Thermodynamics laws}\label{sec5}

The reason behind the exists of non-equilibrium term $d_iS$ in entropy
production is $\rho_d$ and $p_d$ described in eqs. (\ref{2.12}) and
(\ref{2.13}) satisfy the continuity Eq. (\ref{2.17}) whose R.H.S. will not
vanish in $f(R,Y,\phi)$ gravity because $\partial_{t}f_R\neq0$. Otherwise the
standard continuity equation does not satisfy. Now we are again defining the
energy density and pressure of dark components for which the continuity
equation holds and no extra entropy production term occur, which is mentioned
as the equilibrium description. Here in this section we will examine the
equilibrium description of thermodynamics in $f(R,Y,\phi)$ gravity. The field
equation (\ref{2.2}) is redefined as
\begin{eqnarray}\label{4.0}
&&\nonumber G_{\mu\nu}+f_{R}R_{\mu\nu}-\frac{1}{2}\left(f+\omega(\phi) \phi_{;\alpha}
\phi^{;\alpha}\right)g_{\mu\nu}-f_{R;\mu\nu}+g_{\mu\nu}\Box f_{R}+2f_{Y}
R_{\mu}^{\alpha}R_{\alpha\nu}\\
&&-2[f_{Y}R^{\alpha}_{(\mu}]_{;\nu)\alpha}+\Box [f_{Y}R_{\mu\nu}]+[f_{Y}
R_{\alpha\beta}]^{;\alpha\beta}g_{\mu\nu}+\omega(\phi)\phi_{;\mu}\phi_{;\nu}
=G_{\mu\nu}+\kappa T_{\mu\nu}^{(m)},\label{Fieldeq}
\end{eqnarray}
where $\Box=g^{\mu\nu}\nabla_{\mu}\nabla_{\nu}$, $f_{R}=\partial f/\partial R$
and $f_{Y}=\partial f/\partial Y$. The energy-momentum tensor for a perfect fluid is defined as
\begin{equation}\label{4.01}
T_{\mu\nu}^{(m)}= (\rho_{m} + p_{m})u_{\mu}u_{\nu} + p_{m}g_{\mu\nu},
\end{equation}
where $p_{m}$, $\rho_{m}$ and $u_{\mu}$ are the pressure, energy density and
the four velocity of the fluid respectively. Hereafter, we will assume that
the matter of the universe has zero pressure $p_{m}=0$ (dust). The Einstein
field equation from Eq.~(\ref{4.0}) can be written as
\begin{equation}\label{4.02}
R_{\mu\nu}-\frac{1}{2}Rg_{\mu\nu}=8\pi G T_{\mu\nu}^{(m)}+T_{\mu\nu}^{(d)},
\end{equation}
where
\begin{eqnarray}\label{4.03}
\nonumber T_{\mu\nu}^{(d)}&=&-\frac{1}{2}R g_{\mu\nu}f_{R}
+\frac{1}{2}\left(f+\omega(\phi)\phi_{;\alpha}\phi^{;\alpha}\right)g_{\mu\nu}
+f_{R}{}_{;\mu\nu}-g_{\mu\nu}\Box f_{R}-2f_{Y}\\
\nonumber&\times&R_{\mu}^{\alpha}R_{\alpha\nu}+2[f_{Y}R^{\alpha}_{(\mu}]_{;\nu)\alpha}
-\Box[f_{Y}R_{\mu\nu}]-[f_{Y}R_{\alpha\beta}]^{;\alpha\beta}g_{\mu\nu}-\omega
(\phi)\phi_{;\mu}\phi_{;\nu}\\
&+&\left(1-f_R\right)G_{\mu\nu},
\end{eqnarray}
represents an effective energy-momentum tensor related with all the new terms
of the theory.

\subsection{First Law of Thermodynamics}

We are redefining (\ref{2.10}) and (\ref{2.11}) in the form
\begin{eqnarray}\label{4.1}
3\left(H^{2}+\frac{k}{a^{2}}\right)&=&8\pi G(\rho_{m}+\rho_d)\,,\\\label{4.2}
-2\left(\dot{H}-\frac{k}{a^{2}}\right)&=&8\pi G(\rho_{m}+\rho_d+p_d)\,,
\end{eqnarray}
where $\rho_d$ and $p_d$ are the energy density and
pressure of dark components redefined as
\begin{eqnarray}\label{4.3}
\nonumber\rho_d&=&\frac{1}{8\pi G}\bigg[\frac{1}{2}\left(R f_{R}-f\right)
-\frac{1}{2}\omega(\phi)\dot{\phi}^{2}-3H\partial_{t}f_{R}-6H\big(2\dot{H}
+3H^{2}+\frac{k}{a^{2}}\big)\partial_{t}f_{Y}\\
\nonumber&-&f_{Y}\bigg(\dddot{H}+4H\ddot{H}+6\dot{H}H^{2}-2H^{4}-\frac{4kH^{2}}
{a^{2}}\bigg)+3\left(1-f_R\right)\left(H^{2}+\frac{k}{a^{2}}\right)\bigg]\,,\\\\\label{4.4}
\nonumber p_d&=&\frac{1}{8\pi G}\bigg[\frac{1}{2}\left(f-R f_{R}\right)
-\frac{1}{2}\omega(\phi)\dot{\phi}^{2}+\partial_{tt}f_{R}+2H\partial_{t}f_{R}
+\left(4\dot{H}+6H^{2}\right.\\
\nonumber&+&\left.\frac{2k}{a^{2}}\right)\partial_{tt}f_{Y}+4H\bigg(\dot{H}
+3H^{2}+\frac{2k}{a^{2}}\bigg)\partial_{t}f_{Y}+f_{Y}\bigg(4\dddot{H}
+20H\ddot{H}\\
\nonumber&+&10\dot{H}H^{2}+16\dot{H}^{2}-18H^{4}-\frac{18k\dot{H}}{a^{2}}
-\frac{20kH^{2}}{a^{2}}-\frac{8k^{2}}{a^{4}}\bigg)-\left(1-f_R\right)\\
&\times&\left(2\dot{H}+3H^2+\frac{k}{a^2}\right)\bigg]\,.
\end{eqnarray}
In this representation, eq. (\ref{3.5}) becomes
\begin{equation}\label{4.5}
d\tilde{r}_{A}=4\pi GH\tilde{r}_{A}^{3}(\hat{\rho}_{\rm total}
+\hat{p}_{\rm total})dt\,,
\end{equation}
by using the horizon entropy $\hat{S_{h}}$ of the form
\begin{equation}\label{4.6}
\hat{S_{h}}=\frac{A}{4G}\,,
\end{equation}
differentiating Eq. (\ref{4.6}) and using (\ref{4.5}), we have
\begin{equation}\label{4.7}
\frac{1}{2\pi \tilde{r}_{A}}d\hat{S_{h}}=4\pi\tilde{r}_{A}^{3}
\left(\hat{\rho}_{\rm total}+\hat{p}_{\rm total}\right)Hdt\,,
\end{equation}
multiplying both sides of the above equation by $1-\dot{\tilde{r}}_{A}/(2 H
\tilde{r}_{A})$, we have
\begin{equation}\label{4.8}
T_{h} d\hat{S_{h}}=-4 \pi\tilde{r}_{A}^{3}\left(\hat{\rho}_{\rm total}
+\hat{p}_{\rm total}\right)H dt+2\pi\tilde{r}_{A}^{2}\left(\hat{\rho}_{\rm
total}+\hat{p}_{\rm total}\right)d\tilde{r}_{A}\,.
\end{equation}
By defining the Misner-Sharp energy as
\begin{equation}\label{4.9}
\hat{E}=\frac{\tilde{r}_{A}}{2G}=V\hat{\rho}_{\rm total}\,,
\end{equation}
we get
\begin{equation}\label{4.10}
d\hat{E}=4\pi\tilde{r}_{A}^{2}\hat{\rho}_{\rm total}d\tilde{r}_{A}-4\pi
\tilde{r}_{A}^{3}\left(\hat{\rho}_{\rm total}+\hat{p}_{\rm total}\right)Hdt\,.
\end{equation}
Using Eq. (\ref{4.10}) in (\ref{4.8}), we get
\begin{equation}\label{4.11}
T_{h}d\hat{S}_{h}=d\hat{E}-\hat{W}dV\,,
\end{equation}
where we have used the work density $\hat{W}=(1/2)(\hat{\rho}_{\rm total}
-\hat{p}_{\rm total})$ \cite{43}. The equilibrium description of
thermodynamics can be derived by redefining the energy density $\rho_d$ and
the pressure $p_d$ to satisfy the continuity equation.

\subsection{Generalized Second Law of Thermodynamics}

To analyze the equilibrium description of second law of thermodynamics, we
can write the Gibbs equation in terms of all matter and dark energy fluid as
\begin{equation}\label{4.12}
T_{\nu}d\hat{S_{\nu}}=d(\hat{\rho}_{\rm total}V)+\hat{p}_{\rm total}dV\,,
\end{equation}
where $T_{\nu}$ denotes the temperature within the horizon. The second law of
thermodynamics can expressed as
\begin{equation}\label{4.13}
\dot{\hat{S_{h}}}+\dot{\hat{S}}_{\nu}\geq0\,,
\end{equation}
where $\hat{S_{h}}$, $\hat{S}_{\nu}$ are the horizon entropy and the entropy
due to energy sources inside the horizon respectively. Now, we will assume a
relation between the temperature within the horizon and the temperature of
the apparent horizon given by
\begin{equation}\label{4.14}
T_{\nu}=T_{h}\,.
\end{equation}
By substituting Eqs. (\ref{4.11}) and (\ref{4.12}) in Eq.
(\ref{4.13}), we obtain
\begin{equation}\label{4.15}
\dot{S}_{tot}=\dot{\hat{S_{h}}}+\dot{\hat{S_{\nu}}}=\frac{2\pi\Sigma}
{\tilde{r}_{A}R}\geq0\,,
\end{equation}
where
\begin{equation}
\nonumber\Sigma=\frac{1}{2}(\hat{\rho}_{\rm total}+\hat{p}_{\rm total})\dot{V}\,,
\end{equation}
which is the general condition to satisfy the GSLT. Using Eqs. (\ref{4.1})
and (\ref{4.2}), the condition (\ref{4.15}) reduced to
\begin{equation}\label{4.16}
\frac{2\pi H\left(\frac{2k\dot{H}}{a^2}-
\dot{H}^2-\frac{k^2}{a^4}\right)}{G\left(\dot{H}+2H^{2}+\frac{k}
{a^{2}}\right)\left(H^2+\frac{k}{a^2}\right)^{2}}\geq0,
\end{equation}
In case of flat FRW universe to protect the GSLT, the condition (\ref{4.16})
must be satisfied.

\section{Conclusions}\label{sec6}

Scalar tensor theories of gravity appeared as one of the significant
representation in the bunch of alternatives theories of gravity. These
theories proved to be much promising due to their vast applications in
gravitation and cosmology. These theories play key role in developing models
of inflation and DE. In the present paper, we have discussed the
thermodynamical laws in the context of modified
$f(R,R_{\mu\nu}R^{\mu\nu},\phi)$ theory, which involves the Ricci scalar, the
contraction of the Ricci tensor $Y=R_{\mu\nu}R^{\mu\nu}$ (Ricci invariant)
and a scalar field $\phi$. This theory can be regarded as an extended form of
$f(R,\phi)$ gravity. Here, we have presented the general formalism of field
equations for FRW spacetime with any spatial curvature in this theory and
shown that these equations can be cast to the form of FLT
$T_hd\hat{S}_h+T_{in}d\hat{S}_{in}=\delta{Q}$, in non-equilibrium and
$T_hd\hat{S}_h=\delta{Q}$ in equilibrium description of thermodynamics. In
this structure of FLT we find that entropy is contributed from two factors,
first one corresponds to horizon entropy defined in terms of area and second
represents the entropy production term $d\bar{S}$ which is produced because
of the non-equilibrium description in $f(R,Y,\phi)$ gravity. It is worth
mentioning that no such term is present in Einstein, Gauss-Bonnet,
scalar-tensor theory with non-minimally derivative coupling, Lovelock and
braneworld modified theories \cite{24}-\cite{281}. However, in case of modified theories
like $f(R)$ and scalar tensor theories people have suggested various schemes
to avoid the additional entropy term in first law of thermodynamics
\cite{8**,29,30}. Following such approach one can also discuss the
equilibrium thermodynamics as done in \cite{32}.

Moreover, the validity of GSLT at the apparent horizon of FRW
universe is also tested in this modified theory. We present the
general relation involving contributions from horizon entropy,
auxiliary entropy terms and associated with the matter contents
within the horizon is presented in comprehensive way. Here, we have
assumed the proportionality relation between the temperatures
related to apparent horizon and matter components inside the
horizon. To discuss the validity of GSLT, we have selected the more
generic models reconstructed in \cite{34a}, where we consider the de
Sitter and power law cosmological backgrounds. The discussion are
indetail as we have also considered some well known models from
different backgrounds to validate the GSLT. We can retrieve the
results in other modified theories depending on the choice of the
Lagrangian $f(R,Y,\phi)$. When we consider function independent of
$Y$ we can reduce the results of $f(R,Y,\phi)$ into $f(R,\phi)$
gravity and choosing $f(R,Y,\phi)=R\phi$ we get the results for
Brans-Dick theory. Further, by considering a function independent of $Y$
and $\phi$ we get the results for $f(R)$ gravity. Table \ref{Table1}
summarizes the regions where the GSLT is satisfied for all the
models that we discussed in this paper.

In a de-Sitter $f(R,Y,\phi)$ model, one can notice that the validity
of the GSLT depends on five parameters $\alpha_{1}$, $\alpha_{2}$,
$\alpha_{3}$, $\beta$ and $t$ and hence we fixed two parameters,
$\alpha_{1}$ and $\alpha_{2}$ to show the viable regions by varying
the other parameters. Next we have considered $f(R,\phi)$ using
de-Sitter model whose GSLT constraint depends on four parameters
$\alpha_{1}$, $\alpha_{2}$, $\beta$ and $t$. Here we are fixing
$\beta$ and observe the feasible regions by varying the other parameters.
In power law $f(R,\phi)$ case by varying $\alpha_{1}$, $\alpha_{2}$ we
have examined the feasible constraints on $\beta$, $n$ and $t$. Next we have
considered four known models of $f(R,Y,\phi)$ gravity independent of
$Y$, which are of the form $f(R,\phi)$, $Rf(\phi)$, $\phi f(R)$.
Model-I is depending on four parameters $B$, $n$, $\beta$ and $t$,
we have checked the validity of $\dot{S}_{tot}\geq0$ by varying $B$.
Model-II is a function of four parameters $\beta$, $m$, $n$ and $t$,
by fixing $n$ we will discuss the viability of the GSLT for
different values of $\beta$, $m$ and $t$. In model-III the
constraint is depending on four parameters $n$, $\xi$, $\beta$ and
$t$. By fixing $n>1$ we examined the possible regions for the other
parameters. Next in model-IV we have four parameters $n$, $\alpha$,
$\beta$ and $t$. For $n>1$ we have find the feasible constraints on
other parameters.

\begin{table}[H]
    \centering
    \scriptsize{\begin{tabular}{|l|c|c|} \hline \hline
            Models  &  Variations of parameters   & Validity of $\dot{S}_{tot}\geq 0$    \\ \hline \hline
            & $\alpha_{1}>0$, $\alpha_{2}>0$ and  & $\forall$ $t$ with ($\alpha_{3}\geq0$, $\beta\leq-0.78$) or
($\alpha_{3}\leq0$, $\beta\geq0$) \\
            de-Sitter Model
            & $\alpha_{1}<0$, $\alpha_{2}>0$      &          \\  \cline{2-2}\cline{3-2}
            $f(R,Y,\phi)$
            & $\alpha_{1}>0$, $\alpha_{2}<0$ and  & $\forall$ $\alpha_{3}$, $\beta$ and $t$ \\
            & $\alpha_{1}<0$, $\alpha_{2}<0$      &    \\ \hline \hline
            de-Sitter Model $f(R,\phi)$
            &$\forall$ $\beta$                    & $\forall$ $t$ with ($\alpha_{1}\leq-0.1$, $\forall$ $\alpha_{2}$) or
            ($\alpha_{1}>0$, $\alpha_{2}\geq0$) \\ \hline\hline
            & $\alpha_{1}>0$ with $\alpha_{2}<0$  & $n\geq3$ with ($\beta\leq-35.8$, $t\geq1$) or ($\beta\geq2.81$,
$t\geq0.94$)\\
            & ~~~~~~~~~~~~~~ $\alpha_{2}=0$ & $n>1$, $\beta>0$ or $\beta<0$, $\forall$ $t$ \\
            Power Law Model
            & ~~~~~~~~~~~~~~ $\alpha_{2}>0$ & ($n\geq8.6$, $0<\beta\leq20$, $t\geq0.8$) or ($n\geq12.7$, $-20\leq\beta<0$,
$t\geq0.9$)\\
\cline{2-2}\cline{3-2}
            $f(R,\phi)$
            & $\alpha_{1}<0$ with $\alpha_{2}<0$  & ($n\geq8.6$, $-20\leq\beta<0$, $t\geq0.8$) or ($n\geq12.7$, $0<\beta\leq20$,
$t\geq0.9$)\\
            & ~~~~~~~~~~~~~~ $\alpha_{2}=0$       & $n>1$, $\beta>0$ or $\beta<0$, $\forall$ $t$ \\
            & ~~~~~~~~~~~~~~ $\alpha_{2}>0$       & $n\geq3$ and $\forall$ $t$ with ($\beta\leq-28.1$) or ($\beta\geq35.7$)\\
\hline \hline
            & $B=0$ & $n>1$, $\forall$ $\beta$, $t\geq0.96$ \\
            Model-I    & $B>0$ & ($n>1$ with $\beta\leq-0.6$, $\forall$ $t$) or ($n\geq2.5$ with $\beta\geq6$, $t\geq2.5$) \\
            & $B<0$ & not valid                             \\ \hline
            Model-II   & $n>1$ & $\forall$ $t$ with ($m\geq2$, $\beta\leq-1.5$) or ($m\leq-3.2$, $\beta\geq5$) \\ \hline
            Model-III  & $n>1$ & ($\forall$ $\xi$, $\beta\leq-3.5$ and $t\geq4$) or ($\xi\leq0$, $\beta\geq0.15$ and $t\geq1$)
            \\
\hline
            Model-IV   & $n>1$ & ($\beta\geq0$, $\alpha<0$ and $\forall$ $t$) or ($\beta\leq-0.25$, $\alpha\geq0$ and $t\geq1$)
\\ \hline
        \end{tabular}}
        \caption{Validity regions of $\dot{S}_{tot}\geq 0$ for different models.} \label{Table1}
    \end{table}

\vspace{.25cm}

\section{Acknowledgment}

\normalsize \normalfont S.B. is supported by the Comisi\'on Nacional de
Investigaci\'on Cient\'ifica y Tecnol\'ogica (Becas Chile Grant No. 72150066).

\end{document}